# Generative AI as a Tool or Leader? Exploring AI-Augmented Thinking in Student Programming Tasks

Tianlong Zhong[1], Gaoxia Zhu[1]*, Kang You Lim[1], Yew Soon Ong[1]

[1] Nanyang Technological University, 50 Nanyang Ave, Singapore 639798

Tianlong Zhong

Ph.D. Candidate

Graduate College

Nanyang Technological University, Singapore, SINGAPORE

E-mail: tianlong001@e.ntu.edu.sg

Gaoxia Zhu

Assistant Professor

National Institute of Education (NIE)

Nanyang Technological University, Singapore, SINGAPORE

E-mail: gaoxia.zhu@nie.edu.sg

Kang You Lim

Undergraduate Student

College of Computing & Data Science

Nanyang Technological University

E-mail: klim145@e.ntu.edu.sg

Yew Soon Ong

President's Chair Professor of Computer Science

College of Computing & Data Science

Nanyang Technological University, Singapore, SINGAPORE

Email: asysong@ntu.edu.sg

**\* Corresponding author**

Learning Sciences and Assessment Department (LSA), National Institute of Education (NIE),

Nanyang Technological University (NTU), 1 Nanyang Walk, Singapore 637616

E-mail: gaoxia.zhu@nie.edu.sg

Telephone: +65 65928215, Fax: +65 65928215

**Statements and Declarations**

**Competing interests**: The authors declare no potential conflict of interest in the work.

**Availability of data and materials**: Because of confidentiality agreements and ethical concerns, the data used in this study will not be made public. These data will be made available to other researchers on a case-by-case basis.

**Funding**: This study was supported by the Ministry of Education's (MOE) Academic Research Fund (AcRF) (No. RG62/24).

**Acknowledgements**: The authors are indebted to the students who participated in this study.

## Abstract
The increasing use of Generative Artificial Intelligence (GAI) tools in education highlights the need to understand their influence on individuals' thinking processes and agency. This research explored 20 university students' interaction with GAI during programming. Participants completed surveys, recorded their screens during an hour-long programming session, and reflected on their GAI use. To analyse the data, we developed an AI-augmented thinking coding scheme with four dimensions: Question Formulation, Solution Development, Solution Analysis and Evaluation, and Solution Refinement. Participants were categorised into human-led and AI-led groups based on the time ratio of human-generating source code versus copying source code from GAI. T-tests indicated that the human-led group spent significantly more time on Solution Development and Solution Refinement than the AI-led group ($p \leq .05$). Sequential pattern mining revealed distinct patterns of the two groups: the human-led group often refined GAI outputs, while the AI-led group frequently relied on direct answers from GAI. Correlation analyses found that positive attitudes towards AI, critical thinking, and programming self-efficacy positively correlated with Question Formulation; critical thinking was positively related to Solution Refinement; and programming self-efficacy was negatively associated with Solution Analysis and Evaluation. This study enhances understanding of the thinking process in GAI-supported programming.

**Practitioner notes:**
*What is already known about this topic*
· Artificial intelligence and human intelligence each possess unique strengths, and they need to complement one another to achieve greater outcomes.
· ChatGPT and other generative AI tools can help solve programming tasks, provide code explanations, and offer automatic assessments and feedback.
· ChatGPT and other generative AI tools can improve computational thinking, motivation, and programming self-efficacy.

*What this paper adds*
· Human thinking processes augmented by AI can be categorised into four stages: Question Formulation, Solution Development, Solution Analysis and Evaluation, and Solution Refinement.
· Human-led students spend significantly longer time on Solution Development and Solution Refinement than AI-led students.
· AI-led students frequently relied on AI for direct answers, while human-led students critically evaluate and refine AI's output.
· Programming self-efficacy, critical thinking, and attitudes towards AI are associated with AI-augmented thinking.

*Implications for practice*
· AI-augmented thinking framework and codebook provide a structured way to understand how GAI contributes to human thinking processes.
· Students should be at the centre of the thinking process in programming supported by AI. Otherwise, students will be led by AI and lose control of the learning process.
· Educators should cultivate students' programming self-efficacy, critical thinking, and

attitudes towards AI. These factors have the potential to influence the prompt quality and students' ability to effectively evaluate and refine GAI's responses.

**INTRODUCTION**

Generative AI (GAI) technologies like ChatGPT are reshaping education by augmenting teaching practices (e.g., providing instructional materials and assessments), fostering self-directed learning, and advancing learning analytics (Lo, 2023; Yan, Sha, et al., 2024; Yu, 2024; Zhu et al., 2023). However, GAI tools also present challenges, including academic integrity, potential overreliance on AI, and inherent bias (Grassini, 2023; Rahman & Watanobe, 2023). To fully leverage GAI's potential in education, it is crucial for learners to develop a mindset that GAI should serve as a tool to augment their thinking processes rather than a substitute for thinking. However, research on the impact of GAI tools on students' thinking processes is rare.

In the context of programming education, GAI tools have shown the potential to enhance students' computational thinking by assisting with programming tasks and delivering personalised guidance (Wieser et al., 2023; Yilmaz & Karaoglan Yilmaz, 2023b). Nevertheless, their specific influence on learners' thinking processes remains unclear (e.g., Ali et al., 2023; Yilmaz & Karaoglan Yilmaz, 2023a). Programming education emphasizes the development of skills and knowledge required for writing and debugging computer code, which plays a vital role in cultivating problem-solving abilities, logical reasoning, and technological literacy (Agbo et al., 2019). Sun, Boudouaia, Yang, and Xu (2024) compared two approaches to programming with GAI: prompt-based learning, where students received training on crafting effective prompts, versus unprompted learning, where students used simple, everyday language to interact with ChatGPT. Their findings revealed that students in the prompt-based learning condition obtained more precise feedback than those in the unprompted learning condition. Another research by Sun, Boudouaia, Zhu, and Li (2024) found no significant difference in programming performance between students using ChatGPT and those programming without it. These findings raise questions about why the more precise feedback offered by ChatGPT did not translate into improved programming performance. This calls for further investigation into how students interact with GAI tools during programming tasks and how their thinking processes are shaped and reflected by these interactions.

The idea of augmenting human intelligence with artificial mechanisms dates back to the early days of cognitive science and computer science. Pioneers like Alan Turing speculated about machines that could simulate human thought processes (Turing, 1950). Education theorists like Jean Piaget and Lev Vygotsky also explained human thinking processes psychologically. Piaget (1950) believed thinking is the process that originates from humans' response to the environment, with cognitive psychologists like him analysing human thinking based on observable behaviours. On the other hand, Vygotsky (Vygotsky & Cole, 1981) emphasized the importance of social-cultural elements like language in the development of human thinking. Cultural-historical psychologists, following Vygotsky's approach, tend to explore human thinking through discourses. Researchers have developed new frameworks to label and interpret the cognitive process within AI-enhanced environments, such as hybrid intelligence (Dellermann et al., 2019), augmented intelligence (Sharma, 2019), and hybrid-augmented intelligence (Zheng et al., 2017). These frameworks explore the limitations of humans and AI and their complementary strengths, advocating for achieving AI-augmented human cognition. However, most existing studies on AI-augmented human cognition remain theoretical, with limited empirical studies. While some recent work has begun to analyse human interactions with GAI by examining discourse (e.g., prompts) and task-related behaviours (e.g., Sun, Boudouaia, Zhu, et al., 2024), a substantial gap remains in understanding the mental thinking processes underlying learners when using GAI for learning.

An important factor to consider in human-AI interaction is agency since GAI increasingly obscures the distinctions between Artificial intelligence and human intelligence (Yan, Martinez-Maldonado, et al., 2024). Agency in learning is the ability to take the initiative and direct one's own learning process (Moje & Lewis, 2007). Cukurova (2024) proposed a conceptual framework for AI interaction in education for human competence development (AIED-HCD), which focuses on human agency and control in human-AI hybrid intelligence systems. The AIED-HCD framework calls for high human agency in future education. Based on the levels of students' agency in their collaboration with AI, Farrow (2022) classified five types of human-AI collaboration: human centric, human lead, human and AI cooperative, AI lead, and AI centric. Zhu et al. (2024) identified three types of human-generative AI collaboration: human leads, even contribution, and AI leads. They found that 15.19% of participants thought that ChatGPT led the problem-solving process in an interdisciplinary course. These results raise questions about whether and how students with different interaction patterns with AI (human-led versus AI-led) differ in their thinking processes in the programming augmented by AI. Furthermore, factors such as attitude towards AI, critical thinking, and programming self-efficacy may influence students' learning experiences in AI-augmented environments (e.g. Farrow, 2021; Yan et al., 2024; Yilmaz & Karaoglan Yilmaz, 2023b). Nevertheless, it remains unclear how these factors may be associated with the thinking process of students with different interaction patterns with AI.

To address the gaps, this study aims to answer the following research questions (RQs):

RQ1: What categorises student thinking processes in programming tasks supported by GAI?

RQ2: How do Human-led and AI-led students differ in their thinking processes in programming tasks supported by GAI?

RQ3: How do different factors (i.e. programming self-efficacy, critical thinking, and attitude towards AI) associate with the thinking processes in programming tasks supported by GAI?

## LITERATURE REVIEW

*Human AI Relationship: AI-Augmented Thinking*

A principal challenge we currently face is "what AI means in relationship to how humans think and act" (De Cremer & Kasparov, 2021). To make good use of AI for education and mitigate negative impacts, researchers need to develop a better understanding of how AI influences students' thinking processes. The synergy between human intelligence and AI, known as hybrid intelligence (Dellermann et al., 2019), can be manifested in two ways: human-augmented AI and AI-augmented human intelligence. The former, more researched and practised, is about AI systems trained and continuously improved by humans; for instance, many AI systems rely on training datasets generated and labelled by humans (Raykar et al., 2010). In contrast, AI-augmented human intelligence is about how AI systems can augment and extend humans' capabilities; for instance, personal intelligent assistants can help broaden individuals' cognitive bandwidth (Jarrahi et al., 2022). The process of how humans and AI interact determines the efficiency and creativity of the partnership, as Kasparov indicated, "Weak human + machine + better process was superior to a strong computer alone and, more remarkably, superior to a strong human + machine + inferior process" (De Cremer & Kasparov, 2021). Therefore, we need to dive into the process of how AI augments the human thinking process.

Many terms that describe the phenomenon of the mutual enhancement of AI and human capabilities have emerged, such as hybrid intelligence, hybrid-augmented intelligence, augmented intelligence, and AI thinking. Hybrid intelligence argues that machines should better understand human reasoning and operation while humans also should be better aware of AI logic (i.e., AI literacy) (Jarrahi et al., 2021). Dellermann et al. (2019) argued that hybrid intelligence involves two systems. System 1 refers to human intuition, which is fast, and flexible, whereas System 2 refers to AI, which is rational, and consistent. The collaboration between these two systems can help humans quickly access existing knowledge.

Hybrid-augmented intelligence is defined as introducing human intelligence into a computer system to collaborate with AI and achieve the result of "1+1>2" through synthesising human perception, cognition, and AI (Pan, 2016; Zheng et al., 2017). Zheng et al. (2017) proposed a framework for hybrid-augmented intelligence to improve the decision ability of AI systems, which incorporates supervised and unsupervised learning, knowledge bases, and human prediction. They argued that hybrid-augmented intelligence should reduce human involvement to a minimum and delegate most of the work to computers, which contradicts augmented intelligence which emphasises human agency and creativity.

Differently, augmented intelligence places humans' critical thinking, judgment, and common sense at the centre of human-AI collaboration while acknowledging machines' strengths in dealing with repetitive tasks (Sadiku & Musa, 2021). Augmented intelligence (intends to enhance human cognitive performance, such as learning and problem-solving performance, through machine computing (Kim et al., 2022; Yilmaz & Karaoglan Yilmaz, 2023a). Machines do not aim to replace humans but to help humans perform jobs more accurately and efficiently. Guided by augmented intelligence, some studies have applied AI to improve undergraduate students' programming skills (Yilmaz & Karaoglan Yilmaz, 2023a). Compared with hybrid intelligence, and hybrid-augmented intelligence, augmented intelligence emphasises human agency, critical thinking, and creativity as empowered by AI and mitigates humans' overreliance on AI and fear of using AI (Sharma, 2019). This emphasis speaks to the challenge of leveraging the affordances of AI while nurturing human creativity (Kim et al., 2022; Sadiku & Musa, 2021).

Zeng (2013) introduced the concept of "AI thinking", focusing on a generic problem-solving approach that integrates deep learning, cognitive computing, and human thinking. AI thinking aims to provide individuals with an essential mindset on modelling, problem-solving, and data analysis in preparation for an AI-driven world. "AI" represents the machine agent, and "thinking" highlights human-in-the-loop reasoning, aiming to integrate human cognitive processes into the interactions with AI systems (How & Hung, 2019). Despite its potential, AI thinking remains relatively unexplored. Zeng (2013) highlighted that AI thinking requires ongoing updates, modifications, and deeper conceptualisation. Although some studies, such as Ali et al. (2023) and Yilmaz & Karaoglan Yilmaz (2023a), have applied experimental approaches to examine the advantages that GAI offers to students, their specific learning and thinking processes with the support of GAI are not fully understood.

From the studies mentioned above, we found that various terms (e.g., hybrid intelligence, hybrid-augmented intelligence, augmented intelligence, and AI thinking) describe human-AI collaboration and their augmentation to each other. Although the terminologies and emphasis differ, they all describe the collaboration between humans and AI and acknowledge their strengths and weaknesses. However, there are several gaps in the existing literature. First, most studies are theoretical, requiring empirical studies to test the effectiveness of these conceptualisations. Then, these studies depict what should happen during human-AI interaction rather than the actual thinking processes during human-AI interaction. How human thinking processes can be augmented by AI has not been explicitly explored.

*AI in Computing Education*

In the field of computing education research, one of the most important topics is exploring how beginner programmers learn in introductory programming courses (Fincher & Robins, 2019). In higher education, introductory programming courses serve as the gateway for undergraduate students to grasp the essentials of computational thinking and problem-solving. However, given its complexity, introductory programming is challenging for novices (Cheng et al., 2024; Fincher & Robins, 2019). Nearly 28% of undergraduate students in the United States and 28.6% in Singapore fail introductory programming courses (Bennedsen & Caspersen, 2019; Elmaleh & Shankararaman, 2017). Without timely feedback or appropriate support, students can easily give up when struggling with programming concepts, syntactic problems, and debugging Sun & Hsu, 2019; Yukselturk & Altiok, 2017). Thus, it is essential to provide novice programmers with timely support and scaffolding.

Nowadays, LLMs such as ChatGPT are increasingly used in programming education. Wieser et al. (2023) explored the role of ChatGPT in introductory programming courses with 120 undergraduate students. They found that ChatGPT could solve programming tasks, assess programs, and provide individualised instruction and guidelines by analysing the errors in students' code. LLMs have also shown the potential to bridge the programming knowledge gap for students without strong programming backgrounds. For instance, OpenAI Codex, a system built on GPT-3 that can translate the natural language to code, can generate high-quality programming exercises, code explanations, and error explanations (Leinonen et al., 2023; Sarsa et al., 2022).

Computational thinking (CT) is a systematic and algorithmic approach to solving programming problems and is emerging as a general and fundamental thinking and problem-solving skill (Wing, 2006). Multiple frameworks have been proposed to examine the components of CT (e.g., Grover & Pea, 2013; Shute et al., 2017; Wing, 2006). Among these frameworks, four elements are usually mentioned: problem decomposition (i.e., dividing the problem into smaller, more manageable parts), pattern recognition (i.e., identifying recurring similarities in problem-solving processes), abstraction (i.e., simplifying complex problems or systems to their fundamental elements), and algorithms (i.e., constructing a sequential and replicable process to solve problems). Novice programmers face various challenges in CT practices (Medeiros et al., 2019). They may have difficulty describing a programming problem because of the lack of problem-solving skills and the abstract nature of programming. They might encounter difficulties with programming language syntax and structures of control, data, and code. Furthermore, they might encounter difficulties in debugging and tracing the execution.

GAI has the potential to address these challenges. Yilmaz & Karaoglan Yilmaz (2023b) suggested that students who applied GAI tools would encounter less difficulty in syntax problems and spend more time asking original questions (prompts) and practising CT skills. Moreover, programming stands out from other disciplines due to integrated development environments' (IDEs) unique ability to provide immediate feedback. This allows students to continuously debug, learn from their mistakes, and adjust their strategies and behaviours accordingly. Despite the prevalent quasi-experiments on the effects of LLM, little research has delved into how students interact with GAI tools and how the tools play roles in students' thinking processes.

## MATERIALS AND METHODS

*Participants*

To address these research questions, in the January Semester of 2024, 20 students from a university in Singapore were recruited. To ensure a varied participant group, half of the students (4 males and 6 females) were recruited from an undergraduate course on Engineering Computation for an engineering program, while the other half (1 male and 9 females) were from a master's course on Learning Analytics for an education program. All participants' previous programming experience was limited to completing a single module in their program.

*Research Context and Procedures*

The study was conducted out of class during which they performed individual programming activities with the support of GAI tools like ChatGPT. The participants first completed a pre-survey on general attitudes towards AI, critical thinking, and programming self-efficacy (see Instruments for details). They then took part in a one-hour programming session. They recorded their screen during the programming process and were required to reflect on their experience using GAI afterwards. For the undergraduate students, the programming activity was designed by the research team and adapted from Exercism (https://exercism.org/). Students were required to accomplish tasks such as linear search, bubble sorting, class creation, and a text-based user interface using Python. For the master students, the programming activity was an assignment about data analysis using R. They were free to select and analyse data to answer an educational research question they were interested in. Participants were required to record their behaviours in programming IDE and discourses with GAI tools. Figure 1 shows screenshots from a participant's screen recording. Figure 1a shows the programming IDE while Figure 1b presents the learner's interactions in the GAI tool.

**Figure 1**
An example of the programming process supported by GAI tools
1a Programming IDE

1b GAI tool

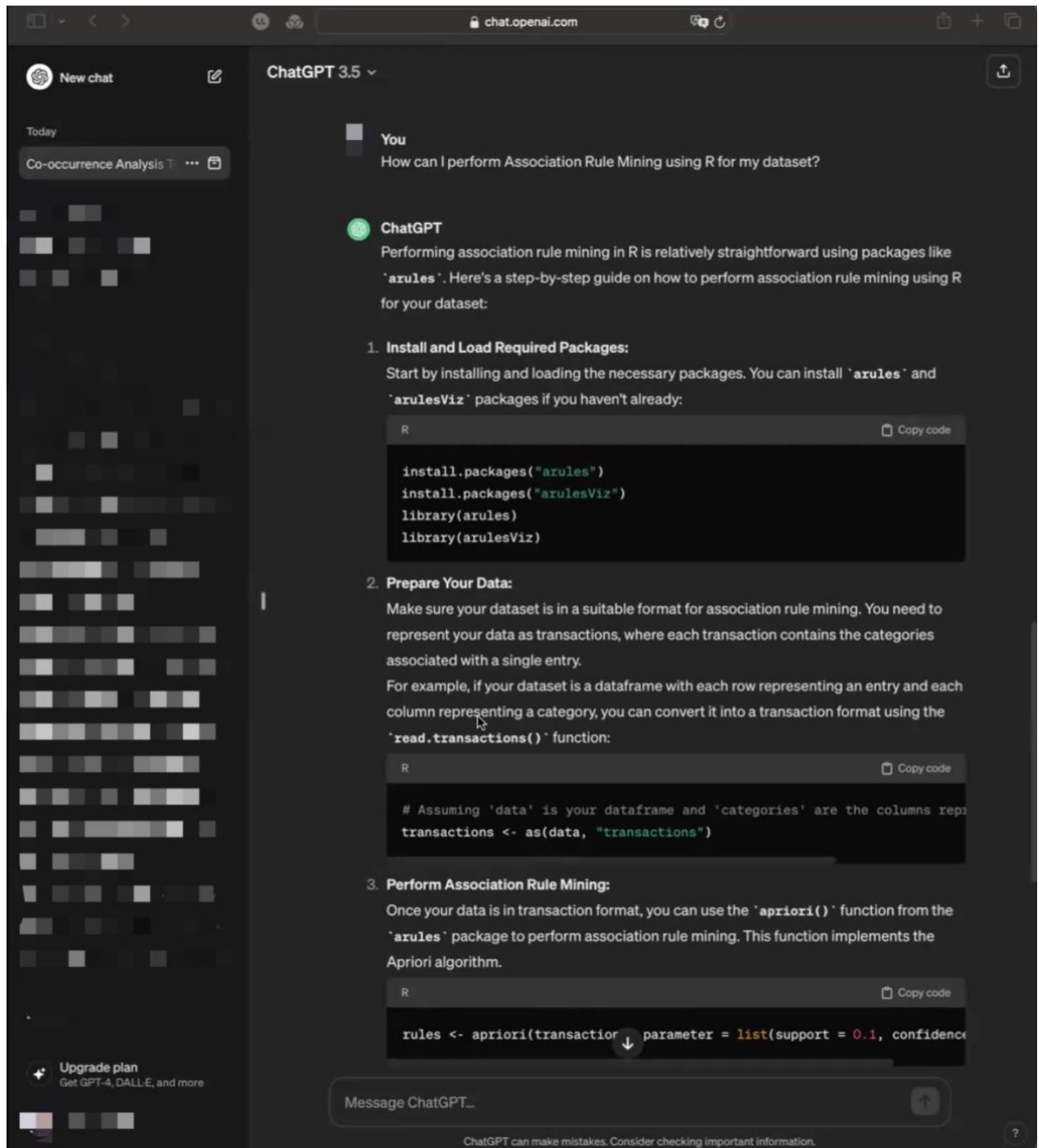

After the programming session, the participants submitted the chat link with GAI tools and filled out the reflection report, which includes the following questions: I) Can you describe any key interactions or moments where GAI significantly influenced the direction or outcome of your task? II) Did GAI work as you expected? Why or why not? III) Were there any challenges or limitations you experienced while using GAI? IV) How likely will you use GAI for similar tasks in the future? Why?

*Instrument*

The following scales were included in the pre-survey that participants filled out because of their high reliability.

I) General attitudes towards AI. We adopted the General Attitudes towards Artificial Intelligence Scale (Schepman & Rodway, 2020). The scale is a five-point Likert scale that contains 20 items. The scale is further divided into two subscales: positive attitudes towards AI (Cronbach's α = 0.88) and negative attitudes towards AI (Cronbach's α = 0.83).

II) Critical thinking. The Critical Thinking Disposition Scale (Sosu, 2013) was adopted in this study. This scale is a five-point Likert scale that contains 11 items (Cronbach's α = 0.81). The scale consists of two subscales: critical openness and reflective skepticism.

III) Programming self-efficacy. The Self-efficacy Scale for Computer Programming (Ramalingam & Wiedenbeck, 1998) was adopted. The scale consists of 32 items, which are further divided into four factors: independence and persistence (Cronbach's α = 0.94), complex programming tasks (Cronbach's α = 0.94), self-regulation (Cronbach's α = 0.86), and simple programming tasks (Cronbach's α = 0.93).

*Data Collection and Analysis*

In the end, all 20 participants submitted the screen recordings (17.98 hours in total) and pre-surveys. Sixteen of them provided the reflection reports. Their names had been coded with the format of course prefix + ID (e.g., RES1, MSLS10) to protect their privacy.

To answer RQ1, we qualitatively analysed the 20 participants' screen recordings. The first author watched the screen recordings, generated qualitative codes regarding students' behaviour in programming IDE and discourses with GAI, and drafted a coding scheme of students' thinking processes in programming with GAI. The analysis unit was each event captured in these recordings, such as task analysis, programming processes, and discourses with GAI. The coding scheme includes the questions asked, the context in which the questions were asked (e.g., while programming or studying new concepts), and how students evaluated and integrated the feedback from GAI into their work. Using the coding scheme, another researcher coded a one-hour screen recording independently, evaluated their comprehensiveness and appropriateness, and provided suggestions for revising it. Then, the research team iteratively refined the coding scheme through discussion and consensus-building. As a result, the final coding scheme (see Table 1) has four main coding categories: (1) Question Formulation: humans describe their questions to generative AI tools. (2) Solution Development: humans and/or AI generate the solutions. (3) Solution Analysis and Evaluation: humans evaluate AI-generated solutions and find their weaknesses and gaps; (4) Solution Refinement: humans revise the solutions by synthesising human and AI input. Each category includes subcategories that emerge in the programming processes. For instance, participants usually asked GAI for source code in the Question Formulation stage and refined code based on GAI answers in the Solution Refinement stage.

Finally, the two researchers coded six screen recordings (753 events) independently using the final coding scheme and achieved a Cohen's Kappa of 0.78, suggesting their substantial agreement and the coding scheme's reliability. The first author then coded the remaining screen recordings using the final coding framework.

To answer RQ2, we began by dividing the participants into two equal groups: AI-led and human-led. For each participant, we calculated the time ratio of human generating source code versus copying source code from GAI based on analysis of their screen recording. Based on these ratios, we ranked all 20 participants. The 10 participants with the highest proportion of time spent generating source code themselves were assigned to the human-led group, while the remaining 10 participants were placed in the AI-led group. Independent t-tests were conducted to check whether the AI-led and human-led groups had significant differences in the four main coding categories of the scheme. Furthermore, we applied sequential pattern mining, specifically the PrefixSpan algorithm (Pei et al., 2004), to detect common thinking sequences in the two groups based on the coded data. Then, we employed case analyses to further check the thinking sequences of these two groups. We selected two representative cases for each group and scrutinised their coded data. Their reflection reports of using GAI were also examined as triangulation.

To answer RQ3, Pearson correlation analyses were utilised to check the correlation between general attitudes towards AI, critical thinking, and programming self-efficacy and AI-augmented thinking codes. There is no consensus on the minimum sample size for Pearson analysis. Twenty is acceptable, as suggested by studies such as Bujang and Baharum (2017) and Humphreys et al. (2019).

## RESULTS
*AI-augmented Thinking Framework and Codebook*

A total of 2,297 events/codes were identified during the analysis of all the screen recordings. Table 1 provides a detailed breakdown of the frequency and duration associated with each code of the 20 participants. The three most frequently observed events were Running source code, GAI generating answers, and Reading answers from GAI. In terms of time investment, the highest durations were Reading answers from GAI, Human generating source code, and Refining source code manually. While students frequently ran source code and leveraged GAI to generate answers, they spent substantial time interpreting AI-generated responses, coding independently, and refining their work manually. This result highlights the intricate interplay between GAI's assistance and human effort in the programming process.

**Table 1**
*The coding scheme of students' thinking processes in programming with GAI*

| Coding categories | Description | Frequency | Time (seconds) |
|---|---|---|---|
| **Question Formulation** | | | |
| Analysing tasks | Human analyses the requirement of the task | 98 | 5521 |
| Asking for source code | Ask GAI tools to generate source code | 47 | 1177 |
| Clarifying requirement | Clarify to GAI tools the requirement of the task (e.g., specify programming language) | 39 | 1348 |
| Copying task description | Copy the task description directly to GAI tools | 41 | 554 |
| Explaining expected outcome | Explain the expected outcomes to GAI tools | 13 | 986 |
| Explaining struggle | Explain the struggle at hand to GAI tools | 13 | 283 |
| Uploading files | Upload files such as csv files and screenshots of programming outcomes to better explain tasks or ask for more specific help | 5 | 205 |
| Writing own questions from task | Asking original questions about the task to GAI tools | 21 | 1262 |
| **Solution Development** | | | |
| GAI generating answers | GAI generates answers based on prompts | 289 | 5946 |
| Human generating source code | Humans generate source code without the help of GAI | 73 | 7493 |
| Preparing working environment | Prepare the environment (e.g., programming IDE, network) | 20 | 1954 |
| Searching in Internet | Searching solutions or knowledge on the Internet | 14 | 498 |
| **Solution Analysis and Evaluation** | | | |
| Asking GAI for explanations | Ask GAI tools to explain the solution | 81 | 1704 |
| Checking outcomes | Check the outcome after executing the source code | 215 | 5353 |
| Copying source code from GAI | Directly copy and paste source code generated by GAI | 214 | 2698 |
| Reading answers from GAI | Learner reads the answers generated by GAI in GAI tools | 230 | 8466 |
| Reading source code in IDE | Read the source code in IDE to understand or check bugs | 76 | 2657 |
| Running source code | Execute the source code in IDE | 451 | 2440 |
| Writing source code based on GAI's answer | Write source code in IDE based on GAI's answer, do not directly copy or paste source code generated by GAI | 35 | 2519 |
| **Solution Refinement** | | | |
| GAI debugging | Copy bug information to ask GAI for debugging | 77 | 2055 |
| Refining source code based on GAI answers | Refine the source code based on GAI answers | 41 | 1667 |
| Refining source code manually | Refine the source code manually in IDE | 189 | 7341 |
| Writing comments | Write comments in IDE to make the code more readable | 15 | 601 |

Building on content analysis of the screen recordings, an AI-augmented thinking framework was constructed, as illustrated in Figure 1. AI-augmented thinking describes an individual's competency to integrate AI with their own thinking by appropriately using AI, critically evaluating AI-generated answers, and applying AI outputs responsibly. AI-augmented thinking process typically begins with formulating a question, resulting in prompts to GAI tools. Following this, students develop solutions by incorporating GAI-generated outputs to varying degrees. If students generate a solution with GAI, they should actively oversee the Solution Development process to mitigate potential inaccuracies or biases. Subsequently, individuals either analyse the solution manually or use GAI to evaluate the solution, identifying any gaps or weaknesses in the solution. The next step involves revising the solution by synthesising contributions from both human and GAI tools. The AI-augmented thinking process is inherently iterative. Students often experience the process multiple times before arriving at a final desirable solution. It is worth noting that we advocated that Human Thinking should be at the core of AI-augmented thinking. Students need to remain active agents/leaders in the learning process rather than merely being led by AI.

**Figure 1**
*AI-augmented thinking framework*

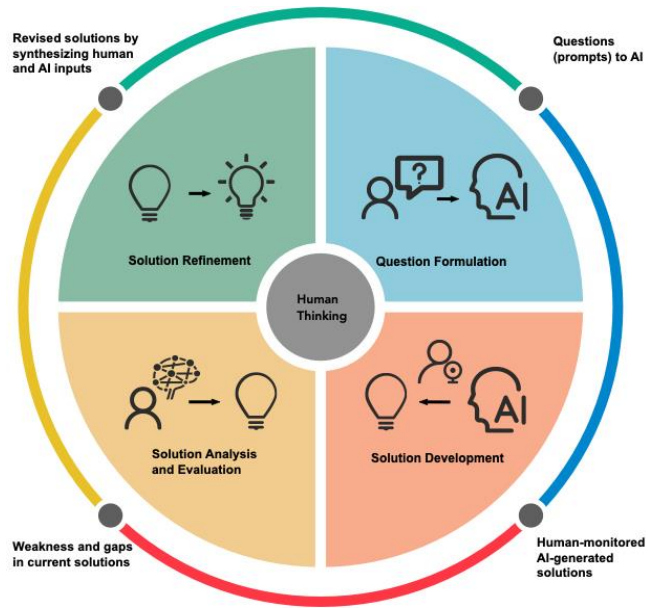

## The Difference in the Thinking Processes Between Human-led and AI-led Students

Regarding RQ2, the results of the t-tests showed no significant difference in Question Formulation, Solution Development, and Solution Analysis and Evaluation. However, Human-led students engaged in Solution Development and Solution Refinement for a longer time than the AI-led students ($p \leq .05$), with moderate power (0.65).

**Table 2**

*Result of t-tests on the time of dimensions between human-led and AI-led students*

| Dimension | T_stat | P_value |
| --- | --- | --- |
| Question Formulation | 0.12 | 0.91 |
| Solution Development | 2.12 | 0.05 |
| Solution Analysis and Evaluation | -1.79 | 0.09 |
| Solution Refinement | 2.48 | 0.03 |

The sequential pattern mining found 25,146 frequent patterns in the coded data. Table 3 shows examples of the most frequent sequences ranked by support (i.e., the percentage of sequences with such subsequences). The result indicated that the human-led participants used GAI as an efficient tool and showed a critical use of GAI to enhance its initial outputs. Common patterns they executed were [Checking outcomes, Reading answers from GAI, Refining source code manually]. They used GAI primarily to automate mundane tasks like generating strings while tackling complex problems independently. They recognised GAI's limitations. RES5 found that "GAI does not have a very holistic view of the solution and only focuses on the current prompt." They strategically integrated it to optimise their workflow. They also had socio-emotional interactions with GAI, as evidenced by their gratitude in chats with GAI, such as "Thank you bestie".

In contrast, AI-led participants had frequent subsequences like [GAI generating answers, Copying source code from GAI] and [Copying source code from GAI, Running source code], suggesting their reliance on GAI for direct answers. They treated GAI like a superintelligence and depended on it to solve programming problems. This reliance tended to reduce their own thinking and manual code refinement. The analyses of the reflection reports suggest similar results, as MSLS2 said: "I think it was very important for GAI to assist me the whole time … I needed to keep asking GAI."

**Table 3**

*Frequent subsequences from data*

| Group | Subsequence | Support |
|---|---|---|
| Human-led | [Checking outcomes, Reading source code in IDE] | 100% |
| | [Checking outcomes, Refining source code manually] | 90% |
| | [Checking outcomes, Reading answers from GAI, Refining source code manually] | 90% |
| | [Reading source code in IDE, Checking outcomes Refining source code manually] | 90% |
| AI-led | [Running source code, Reading answers from GAI, GAI generating answers] | 100% |
| | [GAI generating answers, Copying source code from GAI] | 90% |
| | [Copying source code from GAI, Running source code] | 90% |
| | [GAI generating answers, Copying source code from GAI, GAI generating answers, Copying source code from GAI, Running source code] | 90% |

*Factors associated with thinking processes in programming tasks supported by GAI*

The results of Pearson's correlation analyses are summarised in Table 4. The findings reveal a positive correlation between positive attitudes towards AI and Question Formulation ($r = 0.32$), suggesting students with favourable attitudes towards AI might spend more time crafting questions to ask GAI tools. Critical thinking shows a positive relationship with both Question Formulation ($r = 0.33$) and Solution Refinement ($r = 0.47$), indicating students with stronger critical thinking dispositions tend to dedicate more time to asking questions and refining initial solutions. Additionally, programming self-efficacy is positively associated with Question Formulation ($r = 0.39$) but demonstrates a strong negative correlation ($r = -0.54$) with Solution Analysis and Evaluation, implying that students with greater confidence in programming spent more time crafting prompts but less time analysing and evaluating solutions.

**Table 4**

*Correlation between factors (i.e. programming self-efficacy, critical thinking, and attitude towards AI) and AI-augmented thinking*

| | Question Formulation | Solution Development | Solution Analysis and Evaluation | Solution Refinement |
|---|---|---|---|---|
| Positive attitudes towards AI | 0.32* | -0.29 | -0.18 | 0.22 |
| Negative attitudes towards AI | 0.10 | -0.15 | 0.21 | -0.03 |
| Critical thinking | 0.33* | -0.04 | -0.03 | 0.47* |
| Programming self-efficacy | 0.39* | 0.25 | -0.54** | 0.21 |

\* suggests a moderate correlation
\*\* suggests a strong correlation

## DISCUSSION

This study built a coding scheme of students' thinking processes in programming with GAI and constructed an AI-augmented thinking framework, including four processes: Question Formulation, Solution Development, Solution Analysis and Evaluation, and Solution

Refinement. This work contributes to revisions and further conceptualisation of AI thinking in this GAI age (Zeng, 2013). The descriptive analyses reveal that while students frequently executed code and relied on GAI for solutions, they also dedicated significant time to understanding AI-generated answers, writing source code, and manually refining it. Although GAI serves as a primary resource for navigating programming challenges, its outputs often require substantial human interpretation, analysis, and adaptation. The iterative process of running source code, reviewing AI responses, and making manual adjustments underscores that GAI is not a complete replacement of humans but a complementary and supporting tool in programming tasks, like also in other learning tasks.

We found that human-led students spent significantly more time on Solution Development and Solution Refinement. The result suggests that when students take the lead in directing AI, they are more likely to generate solutions and critically evaluate and refine their solutions, leveraging AI as a supportive tool rather than relying on it passively. The result underscores the importance of promoting student agency in AI-supported learning environments to enhance deep learning and reflective practices, which echoes Cukurov (2024)'s vision for hybrid intelligence, which emphasises the synergistic integration of human and artificial intelligence.

Sequential pattern analysis shows that the human-led participants' common workflow was characterised by sequences like [Checking outcomes, Reading answers from GAI, Refining source code manually], indicating their ability to leverage GAI for routine or repetitive tasks, such as generating strings, while reserving complex, high-level problem-solving for themselves. The patterns demonstrated a critical and strategic use of GAI, treating it as a tool to enhance their productivity while maintaining control over complex problem-solving processes. By strategically integrating GAI to optimise their workflow, these participants not only enhanced their efficiency but also maintained active engagement in the problem-solving process. In contrast, AI-led participants exhibited a more dependent approach, frequently relying on GAI for direct solutions. The recurrent subsequences [GAI generating answers, Copying source code from GAI] and [Copying source code from GAI, Running source code] suggest a pattern of reliance on GAI as a source of ready-made solutions rather than a collaborative tool.

The case study indicates that human-led and AI-led students leverage GAI in different ways. Human-led participants see GAI as a tool for optimising their programming efficiency, whereas AI-led participants often over-rely on it. The result highlights the importance of balancing the agency of humans and AI in learning, avoiding overreliance on AI (Grassini, 2023; Rahman & Watanobe, 2023). This overreliance often diminished students' engagement in deeper cognitive processes, such as manual code refinement or critical evaluation. Fostering students' control over the problem-solving process with GAI can not only enhance reflective practices but also prepare them to critically engage with GAI tools in real-world applications, contributing to the broader goals of hybrid intelligence in education (Cukurov, 2024).

This study found that positive attitudes towards AI, critical thinking, and programming self-efficacy positively correlated with time spent on Question Formulation. The results may be because 1) positive attitudes and confidence in AI might encourage students to engage more actively with AI tools, exploring their capabilities by formulating diverse and thoughtful questions to maximise AI's potential. 2) Question Formulation requires identifying gaps in knowledge and analysing problems from multiple perspectives, skills inherently tied to critical thinking in programming (Wang et al., 2017), which drives deeper inquiry and exploration. 3) Students with higher self-efficacy in programming may feel empowered to ask complex, targeted questions. For example, RES3 argued, "The key was to prompt the GAI carefully, rather than pasting in the task and copy-pasting the resulting code," and effective prompts require programming knowledge and self-efficacy. Our research has extended previous studies (e.g., Yilmaz & Karaoglan Yilmaz, 2023b), which found GAI's impact on improving students' programming skills but did not examine what factors influence GAI's impacts on programming processes.

Additionally, critical thinking was positively related to Solution Refinement. It may be because refining solutions requires students' debugging skills, which have been shown to be highly related to critical thinking (Sun et al., 2024). Students with stronger critical thinking skills are more likely to engage deeply with the debugging process, exploring alternative methods and making informed adjustments to enhance the quality of their solutions. We also found that programming self-efficacy was negatively correlated with Solution Analysis and Evaluation. A possible reason is that, unlike novice programmer who needs a longer time to analyse and evaluate GAI's answers, students with more advanced programming skills may easily find the gaps in current source codes. The correlation analyses suggest the importance of fostering students' positive attitudes towards AI, programming self-efficacy, and critical thinking, as these contribute to crafting high-quality prompts and effectively evaluating and refining GAI's responses.

## IMPLICATIONS AND LIMITATIONS

This study has several limitations. First, the sample size of this study is relatively small. Only 20 students participated in the study. This study uses events as the unit of analysis in the sequential pattern analysis, which increases the sample size to 2,297 events, offering a more granular basis for identifying patterns and trends. Future research can replicate and extend this study with larger sample sizes. Second, it should be noted that we did not test ChatGPT's effectiveness on students' programming performance through experimental design. Instead, we only examined students' thinking processes in a natural setting without intervention, trying to find inherent thinking patterns in programming learning processes supported by GAI. For our next steps, we plan to include pre- and post-tests of students' programming knowledge as well as experimental and comparison conditions to systematically study how GAI influence students' thinking processes and programming performance.

Still, this exploratory but novel study provides theoretical contributions and empirical evidence regarding adopting ChatGPT in programming learning in higher education. Theoretically, we constructed an AI-augmented thinking framework that describes the general thinking process in human-AI interaction, which has the potential to be applied in other contexts to understand the human cognition process. The codebook can be tested and adapted to fit other contexts. For instance, the scheme may be adjusted to analyse students' writing tasks supported by GAI.

Empirically, the results suggest that students cannot directly use GAI's outputs, and their agency is critical in AI-augmented thinking. Students led by AI are very likely to lose control over the programming process and the opportunity to learn programming knowledge, which can lead to further overreliance on AI. Educators should pay attention to the risk that GAI can diminish students' critical thinking and deep understanding/learning (Cukurova, 2024; Dwivedi et al., 2023; Yan, Sha, et al., 2024). The results also suggest the associations between humans' critical thinking, self-efficacy, and attitudes towards AI and their AI-augmented thinking in programming, suggesting the need to monitor, scaffold and instruct students in ways that are appropriate for them.

## CONCLUSION

This study explored college students' thinking processes in programming tasks supported by GAI. We scrutinised 20 participants' behaviours in programming IDE and discourses in GAI. Four important thinking stages emerged from the data: Question Formulation, Solution Development, Solution Analysis and Evaluation, and Solution Refinement. Furthermore, we divided students into two groups based on their ratio of copying source code from GAI and writing their own codes manually. The result of t-tests revealed that Human-led students spent significantly longer time on Solution Refinement. Sequential pattern mining found that the human-led group often followed [Checking outcomes, Reading answers from GAI, Refining source code manually], showing a critical use of AI to enhance initial outputs. In contrast, the AI-led group frequently had common sequences like [Copying source code from GAI, Running source code], suggesting they relied on AI for direct answers. Our findings emphasise the importance of student agency in human-AI interaction. Furthermore, the study finds that programming self-efficacy, critical thinking, and attitudes towards AI are associated with AI-augmented thinking. These factors have the potential to influence the prompt quality and students' ability to effectively evaluate and refine GAI's responses.

Despite its contribution, this study also has limitations, such as its small sample size and limited discipline. Further studies are needed to test AI-augmented thinking frameworks and codebooks with larger sample sizes in different disciplines. Other factors like prior knowledge and motivation should also be considered. Multimodal data such as eye tracking, EEG, and other physiological data can be applied to explore the thinking process in depth. Overall, this research contributes theoretical and empirical insights into AI-augmented thinking processes in programming education.

# REFERENCES

Agbo, F. J., Oyelere, S. S., Suhonen, J., & Adewumi, S. (2019). A Systematic Review of Computational Thinking Approach for Programming Education in Higher Education Institutions. *Proceedings of the 19th Koli Calling International Conference on Computing Education Research*, 1–10. https://doi.org/10.1145/3364510.3364521

Ali, J. K. M., Shamsan, M. A. A., Hezam, T. A., & Mohammed, A. A. Q. (2023). Impact of ChatGPT on Learning Motivation: Teachers and Students' Voices. *Journal of English Studies in Arabia Felix*, *2*(1), Article 1. https://doi.org/10.56540/jesaf.v2i1.51

Bennedsen, J., & Caspersen, M. E. (2019). Failure rates in introductory programming: 12 years later. *ACM Inroads*, *10*(2), 30–36. https://doi.org/10.1145/3324888

Bishara, A. J., & Hittner, J. B. (2012). Testing the significance of a correlation with nonnormal data: Comparison of Pearson, Spearman, transformation, and resampling approaches. *Psychological Methods*, *17*(3), 399–417. https://doi.org/10.1037/a0028087

Bujang, M., & Baharum, N. (2017). A simplified guide to determination of sample size requirements for estimating the value of intraclass correlation coefficient: A review. *The Journal of the School of Dental Sciences, USM*, *12*(1), 1–11.

Cheng, G., Zou, D., Xie, H., & Wang, F. L. (2024). Exploring differences in self-regulated learning strategy use between high- and low-performing students in introductory programming: An analysis of eye-tracking and retrospective think-aloud data from program comprehension. *Computers & Education*, *208*, 104948. https://doi.org/10.1016/j.compedu.2023.104948

Cukurova, M. (2024). The interplay of learning, analytics and artificial intelligence in education: A vision for hybrid intelligence. *British Journal of Educational Technology*, *n/a*(n/a). https://doi.org/10.1111/bjet.13514

De Cremer, D., & Kasparov, G. (2021). AI Should Augment Human Intelligence, Not Replace It. *Harvard Business Review*. https://hbr.org/2021/03/ai-should-augment-human-intelligence-not-replace-it

Dellermann, D., Ebel, P., Söllner, M., & Leimeister, J. M. (2019). Hybrid Intelligence. *Business & Information Systems Engineering*, *61*(5), 637–643. https://doi.org/10.1007/s12599-019-00595-2

Dwivedi, Y. K., Kshetri, N., Hughes, L., Slade, E. L., Jeyaraj, A., Kar, A. K., Baabdullah, A. M., Koohang, A., Raghavan, V., Ahuja, M., Albanna, H., Albashrawi, M. A., Al-Busaidi, A. S., Balakrishnan, J., Barlette, Y., Basu, S., Bose, I., Brooks, L., Buhalis, D., … Wright, R. (2023). Opinion Paper: "So what if ChatGPT wrote it?" Multidisciplinary perspectives on opportunities, challenges and implications of generative conversational AI for research, practice and policy. *International Journal of Information Management*, *71*, 102642. https://doi.org/10.1016/j.ijinfomgt.2023.102642

Elmaleh, J., & Shankararaman, V. (2017). Improving student learning in an introductory programming course using flipped classroom and competency framework. *2017 IEEE Global Engineering Education Conference (EDUCON)*, 49–55. https://doi.org/10.1109/EDUCON.2017.7942823

Farrow, E. (2021). Mindset matters: How mindset affects the ability of staff to anticipate and adapt to Artificial Intelligence (AI) future scenarios in organisational settings. *AI & SOCIETY*, *36*(3), 895–909. https://doi.org/10.1007/s00146-020-01101-z

Farrow, E. (2022). Determining the human to AI workforce ratio – Exploring future organisational scenarios and the implications for anticipatory workforce planning. *Technology in Society*, *68*, 101879. https://doi.org/10.1016/j.techsoc.2022.101879

Fincher, S. A., & Robins, A. V. (2019). *The Cambridge Handbook of Computing Education Research*. Cambridge University Press.

Grassini, S. (2023). Shaping the Future of Education: Exploring the Potential and Consequences of AI and ChatGPT in Educational Settings. *Education Sciences*, *13*(7), Article 7. https://doi.org/10.3390/educsci13070692

Grover, S., & Pea, R. (2013). Computational Thinking in K–12: A Review of the State of the Field. *Educational Researcher*, *42*(1), 38–43. https://doi.org/10.3102/0013189X12463051

How, M.-L., & Hung, W. L. D. (2019). Educing AI-Thinking in Science, Technology, Engineering, Arts, and Mathematics (STEAM) Education. *Education Sciences*, *9*(3), Article 3. https://doi.org/10.3390/educsci9030184

Humphreys, R. K., Puth, M.-T., Neuhäuser, M., & Ruxton, G. D. (2019). Underestimation of Pearson's product moment correlation statistic. *Oecologia*, *189*(1), 1–7. https://doi.org/10.1007/s00442-018-4233-0

Jarrahi, M. H., Lutz, C., & Newlands, G. (2022). Artificial intelligence, human intelligence and hybrid intelligence based on mutual augmentation. *Big Data & Society*, *9*(2), 20539517221142824. https://doi.org/10.1177/20539517221142824

Jarrahi, M. H., Newlands, G., Lee, M. K., Wolf, C. T., Kinder, E., & Sutherland, W. (2021). Algorithmic management in a work context. *Big Data & Society*, *8*(2), 20539517211020332. https://doi.org/10.1177/20539517211020332

Karakose, T., Demirkol, M., Yirci, R., Polat, H., Ozdemir, T. Y., & Tülübaş, T. (2023). A Conversation with ChatGPT about Digital Leadership and Technology Integration: Comparative Analysis Based on Human–AI Collaboration. *Administrative Sciences*, *13*(7), 157. https://doi.org/10.3390/admsci13070157

Kim, J., Davis, T., & Hong, L. (2022). Augmented Intelligence: Enhancing Human Decision Making. In M. V. Albert, L. Lin, M. J. Spector, & L. S. Dunn (Eds.), *Bridging Human Intelligence and Artificial Intelligence* (pp. 151–170). Springer International Publishing. https://doi.org/10.1007/978-3-030-84729-6_10

Leinonen, J., Hellas, A., Sarsa, S., Reeves, B., Denny, P., Prather, J., & Becker, B. A. (2023). Using Large Language Models to Enhance Programming Error Messages. *Proceedings of the 54th ACM Technical Symposium on Computer Science Education V. 1*, 563–569. https://doi.org/10.1145/3545945.3569770

Lo, C. K. (2023). What Is the Impact of ChatGPT on Education? A Rapid Review of the Literature. *Education Sciences*, *13*(4), Article 4. https://doi.org/10.3390/educsci13040410

Medeiros, R. P., Ramalho, G. L., & Falcão, T. P. (2019). A Systematic Literature Review on Teaching and Learning Introductory Programming in Higher Education. *IEEE Transactions on Education*, *62*(2), 77–90. IEEE Transactions on Education. https://doi.org/10.1109/TE.2018.2864133

Moje, E. B., & Lewis, C. (2007). Examining Opportunities to Learn Literacy: The Role of Critical Sociocultural Literacy Research. In *Reframing Sociocultural Research on Literacy*. Routledge.

Pan, Y. (2016). Heading toward Artificial Intelligence 2.0. *Engineering*, *2*(4), 409–413. https://doi.org/10.1016/J.ENG.2016.04.018

Phobun, P., & Vicheanpanya, J. (2010). Adaptive intelligent tutoring systems for e-learning systems. *Procedia - Social and Behavioral Sciences*, *2*(2), 4064–4069. https://doi.org/10.1016/j.sbspro.2010.03.641

Piaget, J. (1950). *The Psychology of Intelligence*. Routledge & Paul.

Rahman, M. M., & Watanobe, Y. (2023). ChatGPT for Education and Research: Opportunities, Threats, and Strategies. *Applied Sciences*, *13*(9), Article 9. https://doi.org/10.3390/app13095783

Ramalingam, V., & Wiedenbeck, S. (1998). Development and Validation of Scores on a Computer Programming Self-Efficacy Scale and Group Analyses of Novice Programmer Self-Efficacy. *Journal of Educational Computing Research*, *19*(4), 367–381. https://doi.org/10.2190/C670-Y3C8-LTJ1-CT3P

Raykar, V. C., Yu, S., Zhao, L. H., Valadez, G. H., Florin, C., Bogoni, L., & Moy, L. (2010). Learning From Crowds. *Journal of Machine Learning Research*, *11*(43), 1297–1322.

Reiche, M., & Leidner, J. L. (2024). Bridging the Programming Skill Gap with ChatGPT: A Machine Learning Project with Business Students. In S. Nowaczyk, P. Biecek, N. C. Chung, M. Vallati, P. Skruch, J. Jaworek-Korjakowska, S. Parkinson, A. Nikitas, M. Atzmüller, T. Kliegr, U. Schmid, S. Bobek, N. Lavrac, M. Peeters, R. van Dierendonck, S. Robben, E. Mercier-Laurent, G. Kayakutlu, M. L. Owoc, … V. Dimitrova (Eds.), *Artificial Intelligence. ECAI 2023 International Workshops* (pp. 439–446). Springer Nature Switzerland. https://doi.org/10.1007/978-3-031-50485-3_42

Sadiku, M. N. O., & Musa, S. M. (2021). Augmented Intelligence. In M. N. O. Sadiku & S. M. Musa (Eds.), *A Primer on Multiple Intelligences* (pp. 191–199). Springer International Publishing. https://doi.org/10.1007/978-3-030-77584-1_15

Sarsa, S., Denny, P., Hellas, A., & Leinonen, J. (2022). Automatic Generation of Programming Exercises and Code Explanations Using Large Language Models. *Proceedings of the 2022 ACM Conference on International Computing Education Research - Volume 1*, *1*, 27–43. https://doi.org/10.1145/3501385.3543957

Schepman, A., & Rodway, P. (2020). Initial validation of the general attitudes towards Artificial Intelligence Scale. *Computers in Human Behavior Reports*, *1*, 100014. https://doi.org/10.1016/j.chbr.2020.100014

Sharma, M. (2019). Augmented Intelligence: A Way for Helping Universities to Make Smarter Decisions. In V. S. Rathore, M. Worring, D. K. Mishra, A. Joshi, & S. Maheshwari (Eds.), *Emerging Trends in Expert Applications and Security* (pp. 89–95). Springer. https://doi.org/10.1007/978-981-13-2285-3_11

Shin, S. (2021). A Study on the Framework Design of Artificial Intelligence Thinking for Artificial Intelligence Education. *International Journal of Information and Education Technology*, *11*(9), 392–397. https://doi.org/10.18178/ijiet.2021.11.9.1540

Shute, V. J., Sun, C., & Asbell-Clarke, J. (2017). Demystifying computational thinking. *Educational Research Review*, *22*, 142–158. https://doi.org/10.1016/j.edurev.2017.09.003

Sosu, E. M. (2013). The development and psychometric validation of a Critical Thinking Disposition Scale. *Thinking Skills and Creativity*, *9*, 107–119. https://doi.org/10.1016/j.tsc.2012.09.002

Sun, C., Yang, S., & Becker, B. (2024). Debugging in Computational Thinking: A Meta-analysis on the Effects of Interventions on Debugging Skills. *Journal of Educational Computing Research*, *62*(4), 1087–1121. https://doi.org/10.1177/07356331241227793

Sun, D., Boudouaia, A., Yang, J., & Xu, J. (2024). Investigating students' programming behaviors, interaction qualities and perceptions through prompt-based learning in ChatGPT. *Humanities and Social Sciences Communications*, *11*(1), 1–14. https://doi.org/10.1057/s41599-024-03991-6

Sun, J. C.-Y., & Hsu, K. Y.-C. (2019). A smart eye-tracking feedback scaffolding approach to improving students' learning self-efficacy and performance in a C programming

course. *Computers in Human Behavior*, *95*, 66–72. https://doi.org/10.1016/j.chb.2019.01.036
Turing, A. (1950). Computing machinery and intelligence. *Mind*, *59*(236), 433.
Vygotsky, L. S., & Cole, M. (1981). *Mind in society: The development of higher psychological processes* (Nachdr.). Harvard Univ. Press.
Wang, X.-M., Hwang, G.-J., Liang, Z.-Y., & Wang, H.-Y. (2017). Enhancing Students' Computer Programming Performances, Critical Thinking Awareness and Attitudes towards Programming: An Online Peer-Assessment Attempt. *Journal of Educational Technology & Society*, *20*(4), 58–68.
Wieser, M., Schöffmann, K., Stefanics, D., Bollin, A., & Pasterk, S. (2023). Investigating the Role of ChatGPT in Supporting Text-Based Programming Education for Students and Teachers. In J.-P. Pellet & G. Parriaux (Eds.), *Informatics in Schools. Beyond Bits and Bytes: Nurturing Informatics Intelligence in Education* (pp. 40–53). Springer Nature Switzerland. https://doi.org/10.1007/978-3-031-44900-0_4
Wing, J. M. (2006). Computational thinking. *Communications of the ACM*, *49*(3), 33–35. https://doi.org/10.1145/1118178.1118215
Yan, L., Martinez-Maldonado, R., & Gasevic, D. (2024). Generative Artificial Intelligence in Learning Analytics: Contextualising Opportunities and Challenges through the Learning Analytics Cycle. *Proceedings of the 14th Learning Analytics and Knowledge Conference*, 101–111. https://doi.org/10.1145/3636555.3636856
Yan, L., Sha, L., Zhao, L., Li, Y., Martinez-Maldonado, R., Chen, G., Li, X., Jin, Y., & Gašević, D. (2024). Practical and ethical challenges of large language models in education: A systematic scoping review. *British Journal of Educational Technology*, *55*(1), 90–112. https://doi.org/10.1111/bjet.13370
Yilmaz, R., & Karaoglan Yilmaz, F. G. (2023a). Augmented intelligence in programming learning: Examining student views on the use of ChatGPT for programming learning. *Computers in Human Behavior: Artificial Humans*, *1*(2), 100005. https://doi.org/10.1016/j.chbah.2023.100005
Yilmaz, R., & Karaoglan Yilmaz, F. G. (2023b). The effect of generative artificial intelligence (AI)-based tool use on students' computational thinking skills, programming self-efficacy and motivation. *Computers and Education: Artificial Intelligence*, *4*, 100147. https://doi.org/10.1016/j.caeai.2023.100147
Yu, H. (2024). The application and challenges of ChatGPT in educational transformation: New demands for teachers' roles. *Heliyon*, *10*(2), e24289. https://doi.org/10.1016/j.heliyon.2024.e24289
Yukselturk, E., & Altiok, S. (2017). An investigation of the effects of programming with Scratch on the preservice IT teachers' self-efficacy perceptions and attitudes towards computer programming. *British Journal of Educational Technology*, *48*(3), 789–801. https://doi.org/10.1111/bjet.12453
Zeng, D. (2013). From Computational Thinking to AI Thinking [A letter from the editor]. *IEEE Intelligent Systems*, *28*(6), 2–4. https://doi.org/10.1109/MIS.2013.141
Zheng, N., Liu, Z., Ren, P., Ma, Y., Chen, S., Yu, S., Xue, J., Chen, B., & Wang, F. (2017). Hybrid-augmented intelligence: Collaboration and cognition. *Frontiers of Information Technology & Electronic Engineering*, *18*(2), 153–179. https://doi.org/10.1631/FITEE.1700053
Zhu, G., Fan, X., Hou, C., Zhong, T., Seow, P., Shen-Hsing, A. C., Rajalingam, P., Yew, L. K., & Poh, T. L. (2023). *Embrace Opportunities and Face Challenges: Using ChatGPT in Undergraduate Students' Collaborative Interdisciplinary Learning* (arXiv:2305.18616). arXiv. https://doi.org/10.48550/arXiv.2305.18616

Zhu, G., Sudarshan, V., Kow, J. F., & Ong, Y. S. (2024). *Human-Generative AI Collaborative Problem Solving Who Leads and How Students Perceive the Interactions* (arXiv:2405.13048). arXiv. https://doi.org/10.48550/arXiv.2405.13048